# Coherence factors in a high-$T_c$ cuprate probed by quasi-particle scattering off vortices

T. Hanaguri[1,2], Y. Kohsaka[1,2], M. Ono[1,2], M. Maltseva[3], P. Coleman[3], I. Yamada[4*], M. Azuma[4], M. Takano[4†], K. Ohishi[5‡], H. Takagi[1,6]

1. Magnetic Materials Laboratory, RIKEN, Wako 351-0198, Japan.
2. CREST, Japan Science and Technology Agency, Kawaguchi 332-0012, Japan.
3. Department of Physics and Astronomy, Rutgers University, Piscataway, NJ 08854-8019, USA.
4. Institute for Chemical Research, Kyoto University, Uji 601-0011, Japan.
5. Advanced Science Research Center, Japan Atomic Energy Agency, Ibaraki 319-1195, Japan.
6. Department of Advanced Materials, University of Tokyo, Kashiwa 277-8561, Japan.

\* present address: Department of Chemistry, Ehime Universiy, Matsuyama 790-8577, Japan.
† present address: Institute for Integrated Cell-Material Sciences, Kyoto University, c/o Research Institute for Production Development, Kyoto 606-0805 Japan.
‡ present address: Advanced Meson Science Laboratory, RIKEN, Wako 351-0198, Japan.

Abstract

Coherence factors are a hallmark of superconductivity as a pair-condensation phenomenon. When electrons pair, quasi-particles develop an acute sensitivity to different types of scattering potential, described by the appearance of coherence factors in the scattering amplitudes. While the effects of coherence factors are well established in isotropic superconductors, they are much harder to detect in their anisotropic counterparts, such as high-$T_c$ cuprates. Here we demonstrate a new approach which highlights the momentum-dependent coherence factors in $Ca_{2-x}Na_xCuO_2Cl_2$. Using Fourier-transform scanning tunnelling spectroscopy to detect quasi-particle interference effects, our experiments reveal a magnetic-field dependence in quasi-particle scattering which is sensitive to the sign of the anisotropic gap. This result can be understood in terms of *d*-wave coherence factors and it exposes the role of vortices as quasi-particle scattering centers. We also show that a magnetic field gives rise to an enlarged gapless region around the gap nodes.

Superconductivity is characterized by macroscopic phase coherence, as exemplified by the development of a complex order parameter. In this respect, superconductors are similar to bosonic superfluids, such as $^4$He, and many phenomenological features are common to both, such as vortex quantization and the Josephson effect (*1*). However, superconductors are microscopically distinct from bosonic superfluids, for whereas bosons individually condense, electrons condense as Cooper pairs.

The internal structure of the condensed pairs strongly influences the properties of quasi-particle excitations, enforcing coherence between two quasi-particle scattering processes, $\mathbf{k}_i \rightarrow \mathbf{k}_f$ and $-\mathbf{k}_f \rightarrow -\mathbf{k}_i$. ($\mathbf{k}_i$ and $\mathbf{k}_f$ denote initial and final state momenta, respectively.) This manifests itself as a coherence factor $C(\mathbf{k}_i,\mathbf{k}_f)$ in the scattering matrix element, which is sensitive to the momentum-dependent phase of the superconducting (SC) order parameter and the time-reversal symmetry of the scattering potential (*2*). Studies of coherence factors should therefore provide insight into the nature of electron pairing and quasi-particle scattering processes in unconventional superconductors.

In *s*-wave superconductors, the effect of coherence factors is manifested in the temperature dependence of various measurable quantities associated with quasi-particle scattering. For example, the nuclear-spin relaxation rate exhibits an enhancement just below the SC transition temperature $T_c$, called the Hebel-Slichter peak (*3*), which is a consequence of coherence factors. In unconventional superconductors, however, the effect of coherence factors is suppressed by the strong momentum $\mathbf{k}$ dependence of the anisotropic gap (*4*).

We propose a new technique to highlight the $\mathbf{k}$-dependent coherence factors in anisotropic superconductors by introducing vortices as controllable scatterers. The vortices induced by a magnetic field $B$ will generate extra quasi-particle scatterings around them, in which coherence factor effects can be detected. Quasi-particle scattering can be examined through the Fourier analysis of quasi-particle interference (QPI) patterns imaged by spectroscopic-imaging scanning tunneling microscopy (SI-STM) (*5-7*). Here we report a field-dependent QPI study in a high-$T_c$ cuprate Ca$_{2-x}$Na$_x$CuO$_2$Cl$_2$. The *B*-dependence and spatial variation of QPI intensities, that we observe, can be naturally understood as a manifestation of the *d*-wave coherence factor and establishes vortices as scattering centers with a particular momentum selectivity. A field-induced change in the SC gap dispersion, which results an enlarged gapless region near the SC gap node, is also revealed by our measurements.

First we review the QPI effect. In general, QPI is produced by elastic scattering which mixes the quasi-particle states along the contour of constant energy in **k** space. This gives rise to standing waves of particular scattering wavevectors $\mathbf{q} \equiv \mathbf{k}_f - \mathbf{k}_i$. To image these standing waves, we use SI-STM to map the tunneling conductance $g(\mathbf{r}, E) \equiv dI/dV(\mathbf{r}, E)$, where *I* and *V* are tunneling current and bias voltage, respectively. $g(\mathbf{r}, E)$ is a measure of the local density of states (DOS) at location **r** and energy *E*. The amplitude and **q** of the standing wave can be accurately determined from the Fourier transform of $g(\mathbf{r},E)$, $g(\mathbf{q},E)$. $g(\mathbf{q},E)$ is proportional to the scattering probability (*8*), given by Fermi's golden rule,

$$g(\mathbf{q}, E) \propto \iint |V(\mathbf{k}_i, \mathbf{k}_f)|^2 J(\mathbf{k}_i, \mathbf{k}_f, E) \delta(\mathbf{q} - (\mathbf{k}_f - \mathbf{k}_i)) \delta(E - E(\mathbf{k}_i)) \delta(E - E(\mathbf{k}_f)) d\mathbf{k}_i d\mathbf{k}_f \quad ,$$

where $V(\mathbf{k}_i, \mathbf{k}_f)$ is a **k**-dependent scattering matrix element and $J(\mathbf{k}_i, \mathbf{k}_f, E)$ denotes a joint DOS. The dispersion of quasi-particles in **k** space, $E(\mathbf{k})$, can be experimentally determined from the *E* dependence of **q**. Indeed, in conventional metals and semiconductors, QPI has been widely used to determine the surface band structures (*9-11*).

QPI patterns in the high-$T_c$ cuprates are dominated by a small set of wavevectors **q** given by the 'octet model' (*6*). The low-energy excitations in the SC states are Bogoliubov quasi-particles with a dispersion $E(\mathbf{k}) = \sqrt{\varepsilon(\mathbf{k})^2 + \Delta(\mathbf{k})^2}$, where $\varepsilon(\mathbf{k})$ and $\Delta(\mathbf{k})$ are dispersion relations of normal-state band and SC gap, respectively (*1, 2*). In high-$T_c$ cuprates, the SC gap has $d_{x^2-y^2}$-wave symmetry (*12*) and vanishes along the (±π, ±π) directions (diagonals of the unit cell). The dispersion $E(\mathbf{k})$ gives rise to four 'banana-shaped' contours of constant energy, as shown in Fig. 1A. The amplitudes of the standing waves become large if the momentum transfer **q** connects the ends of the 'bananas', where the joint DOS is the largest. This determines the locations of seven distinct scattering vectors ($\mathbf{q}_i$ (*i* = 1 ~ 7) in Fig. 1A) in **q** space. The quasi-particle states located at the ends of the bananas lie on the normal state Fermi surface (red curves in Fig.

1A), where $\varepsilon(\mathbf{k}) = 0$, and the energy $E(\mathbf{k}) = |\Delta(\mathbf{k})|$ at these points corresponds to the magnitude of SC gap.

In the SC state, the coherence of quasi-particle scattering induced by pair formation causes the scattering probability to acquire an additional $\mathbf{k}$ dependence determined by the coherence factor $C(\mathbf{k}_i,\mathbf{k}_f)$ (*2*). $C(\mathbf{k}_i,\mathbf{k}_f)$ is given by a combination of Bogoliubov coefficients $u_\mathbf{k} = \text{sgn}(\Delta(\mathbf{k}))\sqrt{(1+\varepsilon(\mathbf{k})/E(\mathbf{k}))/2}$ and $v_\mathbf{k} = \sqrt{1-u_\mathbf{k}^2}$. The detailed form of $C(\mathbf{k}_i,\mathbf{k}_f)$ depends on the nature of the scatterer as summarized in table 1 (*2, 13-17*). For a scalar potential, which is even under time reversal, $C(\mathbf{k}_i,\mathbf{k}_f) = (u_{\mathbf{k}_i}u_{\mathbf{k}_f} - v_{\mathbf{k}_i}v_{\mathbf{k}_f})^2$, for a magnetic scattering potential, which is odd under time reversal, $C(\mathbf{k}_i,\mathbf{k}_f) = (u_{\mathbf{k}_i}u_{\mathbf{k}_f} + v_{\mathbf{k}_i}v_{\mathbf{k}_f})^2$. Scattering off inhomogeneities in the SC gap amplitude, which is a kind of inhomogeneous Andreev reflection process that converts electrons into holes as they are scattered, gives rise to the coherence factor $C(\mathbf{k}_i,\mathbf{k}_f) = (u_{\mathbf{k}_i}v_{\mathbf{k}_f} + v_{\mathbf{k}_i}u_{\mathbf{k}_f})(u_{\mathbf{k}_i}u_{\mathbf{k}_f} + v_{\mathbf{k}_i}v_{\mathbf{k}_f}) \propto \Delta(\mathbf{k}_i) + \Delta(\mathbf{k}_f)$. As shown in Fig 1B and C, $u_\mathbf{k}$ changes its sign in the same way as $\Delta(\mathbf{k})$ while $v_\mathbf{k}$ is always positive. This leads to a systematic '*extinction rule*' for $\mathbf{q}_i$'s which depends on the nature of the scatterer. In the case of weak scalar potential scattering, $C(\mathbf{k}_i,\mathbf{k}_f) \sim 0$ for those $\mathbf{q}_i$ which *preserve* the sign of SC gap $\Delta(\mathbf{k})$, namely, $\mathbf{q}_1$, $\mathbf{q}_4$ and $\mathbf{q}_5$. By contrast, for scattering off magnetic impurities or gap inhomogeneities, $C(\mathbf{k}_i,\mathbf{k}_f) \sim 0$ for those $\mathbf{q}_i$ which *reverse* the sign of $\Delta(\mathbf{k})$. Thus, depending on the type of disorder, sign-reversing, or sign-preserving scatterings will dominate (*13-17*). In this way, QPI patterns can shed light onto the underlying nature of the quasi-particle scattering processes. However, in experiments carried out to date (*5-7*), each of the $\mathbf{q}_i$'s has been featured with comparable intensity, which indicates that more than one kind of scatterers is present in the samples, hiding the underlying effects of $C(\mathbf{k}_i,\mathbf{k}_f)$.

The introduction of vortices by applying a magnetic field $B$ provides the system with scatterers with definite $\mathbf{q}$ selectivity. The phase of the SC gap precesses by $2\pi$ about each vortex, whereas the amplitude of the gap vanishes at its core. Both the phase gradient, proportional to the superfluid velocity, and the inhomogeneity in the SC gap amplitude, induced by the vortex core, can scatter quasi-particles. The inhomogeneous superflow about a vortex produces Doppler-shift scattering (*18*) which is odd under time reversal with $C(\mathbf{k}_i,\mathbf{k}_f) = (u_{\mathbf{k}_i}u_{\mathbf{k}_f} + v_{\mathbf{k}_i}v_{\mathbf{k}_f})^2$ like magnetic impurities. The spatial inhomogeneity in the SC gap amplitude causes inhomogeneous Andreev scattering with $C(\mathbf{k}_i,\mathbf{k}_f) = (u_{\mathbf{k}_i}v_{\mathbf{k}_f} + v_{\mathbf{k}_i}u_{\mathbf{k}_f})(u_{\mathbf{k}_i}u_{\mathbf{k}_f} + v_{\mathbf{k}_i}v_{\mathbf{k}_f})$. It should be noted that both of these scatterers selectively activate the sign-preserving **q** points. Therefore, the effect of $C(\mathbf{k}_i,\mathbf{k}_f)$ and the nature of quasi-particle scattering off vortices can be revealed through the **q**-dependence of the *B*-induced change of QPI.

We performed SI-STM measurements in a field on nearly optimally-doped $Ca_{2-x}Na_xCuO_2Cl_2$ ($x \sim 0.14$, $T_c \sim 28$ K) single crystals (*7, 19*) using a low-temperature ultrahigh-vacuum scanning tunneling microscope. Samples were cleaved *in situ* at 77 K and transferred to the microscope maintained at a temperature below 10 K. In order to make the vortex distribution inside the sample uniform, we applied fields up to 11 T along the *c* axis at 5 K, where we confirmed by magnetization measurements that vortex pinning was negligibly small. Then, the samples were field-cooled down to 1.6 K where all the data were collected. At 1.6 K, pinned vortices can be observed as shown in Fig. 2A, B, and C and the number of vortices is consistent with the expected number. All the spectroscopic data were taken in the same field of view simultaneously with atomic-resolution topographic images.

In $Ca_{2-x}Na_xCuO_2Cl_2$, the raw $g(\mathbf{r},E)$ data are dominated by 'checkerboard' modulations (*20*) which mask the underlying QPI signal. As we reported earlier (*7*), the QPI signal is enhanced by taking the ratio $Z(\mathbf{r},E) \equiv g(\mathbf{r},E)/g(\mathbf{r},-E)$. This procedure almost completely suppresses the checkerboard signal. It has an additional advantage of eliminating extrinsic effects associated with the scanning feedback loop and thus $Z(\mathbf{r},E)$ faithfully represents the local-DOS ratio (*7,8*). Detailed characteristics of $Z(\mathbf{r},E)$ have recently been discussed both theoretically and experimentally (*17, 21*).

Figures 2D and G show the zero-field $Z(\mathbf{r},E = 4.4$ meV$)$ and its Fourier transform $|Z(\mathbf{q},E = 4.4$ meV$)|$, displaying the full set of discrete **q**-points expected in the octet model. These discrete **q**-points disperse with *E* in a fashion consistent with a *d*-wave SC gap up to $10 \sim 15$ meV, which marks an upper limit for the detection of well-defined Bogoliubov quasi-particles (Fig. 5) (*7*).

When we repeated the measurements at $B$ = 5 T and 11 T, we observed a remarkable field dependence in the intensities of the QPI patterns. The $B$ dependence of $Z(\mathbf{r},E)$ is shown in Figs. 2D, E, and F. The corresponding Fourier transformed data, shown in Fig. 2G, H, and I, show that field does not induce additional **q** vectors and the positions of the peaks in $|Z(\mathbf{q},E)|$ are only weakly $B$ dependent. By contrast, there is a significant $B$ dependence in the intensities of the peaks. Depending on $\mathbf{q}_i$, intensity of peak is either enhanced or suppressed.

To explore the details of the field-induced intensity variations, we subtract the zero-field data $|Z(\mathbf{q},E, B = 0)|$ from $|Z(\mathbf{q},E,B)|$. The corresponding difference maps are shown in Figs. 3A for $B$ = 11 T. It is clear that we can classify the **q**-points into two groups: for $\mathbf{q}_1$, $\mathbf{q}_4$, and $\mathbf{q}_5$ the intensity is field-enhanced, while for $\mathbf{q}_2$, $\mathbf{q}_3$, $\mathbf{q}_6$, and $\mathbf{q}_7$ the intensity is depressed. These two groups are nothing but the sign-preserving and sign-reversing **q**-points discussed earlier, and their selective enhancement and suppression imply the activation of coherence factors $C(\mathbf{k}_i,\mathbf{k}_f)$ induced by vortices.

The spatial resolution of SI-STM allows us to spatially resolve the origin of the momentum-selective enhancement and suppression of the QPI signal and thereby, to examine its relationship with the location of the vortices. For this purpose, we restricted the field of view to the vicinity of vortices (as indicated by the 'Vortex region' inside the blue lines in Fig. 3B) or to regions far from vortices (as indicated by the 'Matrix region' inside the red lines in Fig. 3B), and performed Fourier analyses separately for each region (Fig. 6). As shown in Fig. 3C and D, the enhanced sign-preserving QPI signals are concentrated in the 'Vortex region', indicating that this signal is associated with coherent scattering effects induced by vortices. By contrast, the suppressed sign-reversing QPI signals are apparently weak near vortices but are distributed throughout the 'Matrix region' far from the vortices, and may be associated with the superflow surrounding the vortices.

We note as an aside that the enhanced sign-preserving **q** points $\mathbf{q}_1$ and $\mathbf{q}_5$ are very close to those wavevectors which characterize the 'checkerboard' electronic modulation observed in the vicinity of vortices in $Bi_2Sr_2CaCu_2O_y$ (*22,23*). Various charge- and spin-density-wave scenarios have been advanced to account for the vortex 'checkerboard' modulation (*22,23*). In such scenarios, the characteristic wavevector of the scattering is not expected to disperse with energy. However, in our data the field-enhanced QPI intensities at these **q**-points do disperse with energy (Fig. 5), so the electronic-order scenario does not appear to apply, at least, in its simplest form. Further studies of the relation between QPI and the vortex 'checkerboard' are needed to elucidate the electronic structure of vortices in high-$T_c$ cuprates.

The reduction of the scattering at the sign-reversing vectors $q_2$, $q_3$, $q_6$, and $q_7$ in the matrix region can be accounted for in terms of the Doppler shift of quasiparticle energies induced by the superflow around vortices (*18*). The Doppler shift in the quasiparticle energies deforms the banana-shaped contours of constant energy by an amount proportional to the superfluid velocity. This has the effect of smearing the quasiparticle interference peaks, reducing their amplitude. This effect has no momentum selectivity, and will tend to uniformly depress the scattering at all octet momenta. However, in practice, we are unable to completely mask out the strong enhancement effects induced by the vortices, so that the smearing effect is only observed at the sign-reversing momenta. A similar scenario has also been proposed by Pereg-Barnea and Franz (*17*).

Finally, we examine the *B* dependence of Fermi surface topology and SC gap dispersion $\Delta(\mathbf{k})$ by analyzing the *E* dependence of $|Z(\mathbf{q},E,B)|$. As shown in Fig. 4A, the Fermi surface displays no measurable *B* dependence up to 11 T as expected, since the corresponding Zeeman energy to *B* is negligibly small (< 1 meV), compared with the hopping amplitude (~ 0.1 eV). On the other hand, $\Delta(\mathbf{k})$ shows a small but distinct field dependence as shown in Fig. 4B. At *B* = 0 T, a linear extrapolation of $\Delta(\mathbf{k})$ from high *E* does not intercept the node at $\theta_\mathbf{k} = 45°$, where $\theta_\mathbf{k}$ is a Fermi surface angle around $(\pi,\pi)$, and there is an apparent 'gapless' region around the node. This may indicate that $\Delta(\mathbf{k})$ contains a higher order harmonic such as $\cos(6\theta_\mathbf{k})$ (*24*) or, alternatively, that $\Delta(\mathbf{k})$ actually vanishes in a finite region around the node due to pair-breaking scattering (*25*) off impurities. This 'gapless' region expands in a field, indicating that zero-energy quasi-particles are generated by introducing vortices. In accordance with this interpretation, the observed DOS at the

Fermi energy, given by the average value of $g(\mathbf{r}, E = 0)$, increases as shown in Fig. 4C and is seen to follow a $B\log B$ behaviour (*26*), as expected in a dirty *d*-wave superconductor (Fig. 4D). These results provide a spectroscopic basis for the field-induced DOS observed by specific heat (*27*) and nuclear magnetic resonance (*28*) measurements and are consistent with the Volovik effect (*18*) which predicts field-induced gapless excitations around the SC gap nodes in **k** space and outside the vortex core in real space.

In conclusion, we have detected the *d-wave coherence factor* of a high-$T_c$ cuprate using vortices as controllable quasi-particle scattering centers. Our results establish that vortices selectively activate those quasi-particle scattering channels that preserve the sign of the SC gap in **k** space. Future measurements using this technique may offer the opportunity to probe the nature of the anomalous electronic matter inside the vortex core using quasi-particle vortex scattering. Moreover, this method provides a simple phase-sensitive probe of gap anisotropy that can be applied to other superconductors with other forms of anisotropic gap, such as *p*-wave and extended *s*-wave superconductors. Another variant on this method is to examine the coherence factors for scattering off conventional impurities at temperatures above $T_c$ (*29*): this approach may provide a viable way to probe the nature of the order that develops in the pseudogap normal state (*30*). Finally, we note that Fourier-transform SI-STM is currently the only method to study the evolution of **k**-dependent electronic states as a function of magnetic field and, in this respect, offers a useful tool for the study of a wide range of field-induced quantum phenomena (*31*).

32. The authors thank J. C. Davis, H. Eisaki, M. Franz, C. -M. Ho, K. Machida, T. Pereg-Barnea and P. Wahl for valuable discussions and comments. T. H., M. T., and H. T. are supported by Grant-in-Aid for Scientific Research from the Ministry of Education, Culture, Sports, Science and Technology of Japan. M. M. and P. C. are supported by NSF DMR 0605935.

**Table S1**. Coherence factors $C(\mathbf{k}_i,\mathbf{k}_f)$ associated with various scatterers and corresponding enhanced $\mathbf{q}_i$ in QPI patterns.

| Scatterer | Coherence factor | Enhanced $q_i$ |
| --- | --- | --- |
| Weak scalar | $(u_{\mathbf{k}_i}u_{\mathbf{k}_f} - v_{\mathbf{k}_i}v_{\mathbf{k}_f})^2$ | 2, 3, 6, 7 |
| Magnetic, phase-gradient | $(u_{\mathbf{k}_i}u_{\mathbf{k}_f} + v_{\mathbf{k}_i}v_{\mathbf{k}_f})^2$ | 1, 4, 5 |
| Gap amplitude | $(u_{\mathbf{k}_i}v_{\mathbf{k}_f} + v_{\mathbf{k}_i}u_{\mathbf{k}_f})(u_{\mathbf{k}_i}u_{\mathbf{k}_f} + v_{\mathbf{k}_i}v_{\mathbf{k}_f})$ | 1, 4, 5 |

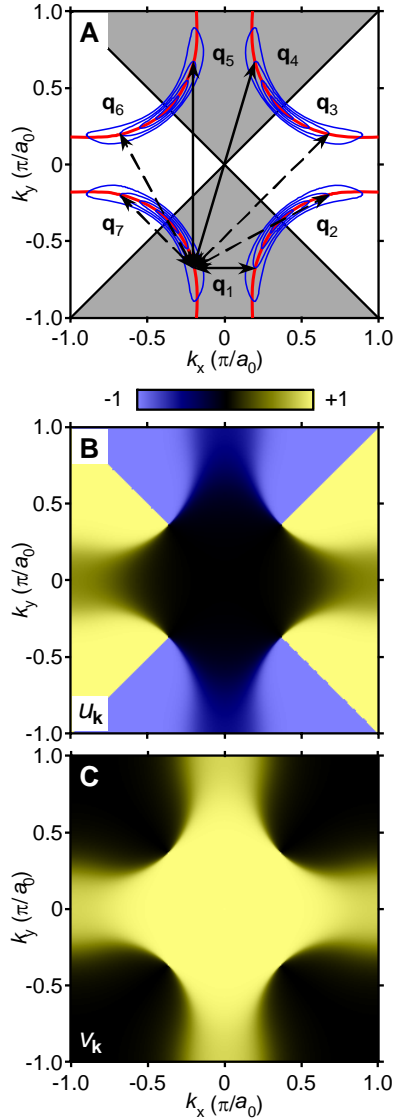

**Fig. 1.** Schematic representation of **k**-space electronic states in a high-$T_c$ cuprate. (**A**) Normal-state Fermi surface (red curves) and contours of constant energy for Bogoliubov quasi-particles (blue curves) in the 1st Brillouin zone. White and shaded areas represent **k**-space regions with opposite signs of $d$-wave SC gap. Arrows denote scattering **q** vectors responsible for QPI patterns. They are classified into sign-preserving and sign-reversing vectors indicated by solid and broken arrows, respectively, according to the relative signs of SC gap between initial and final states. These two kinds of vectors are associated with different coherence factors as summarized in table 1. (**B** and **C**) Bogoliubov coefficients $u_\mathbf{k}$ (B) and $v_\mathbf{k}$ (C) are mapped in **k** space. Note that $u_\mathbf{k}$ changes its sign according to that of SC gap, while $v_\mathbf{k}$ is always positive.

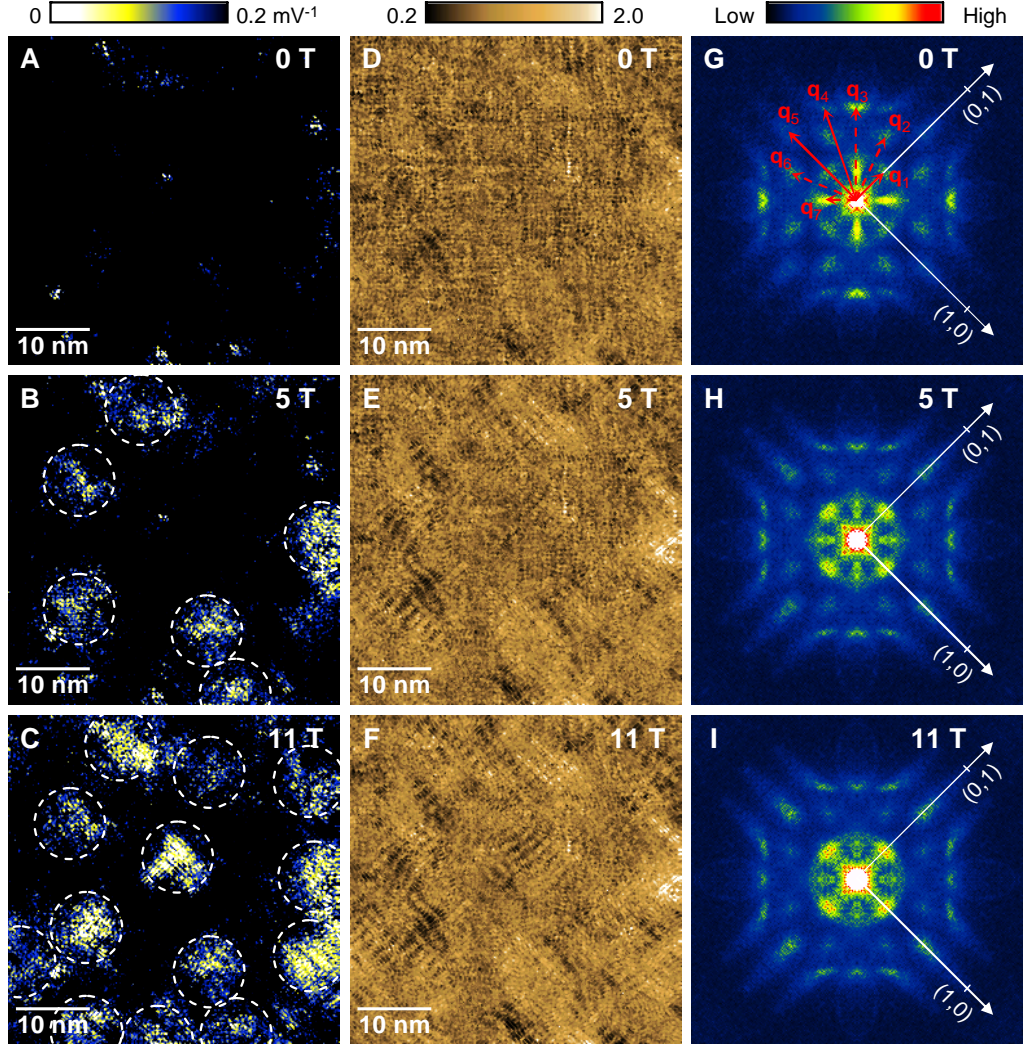

**Fig. 2.** Imaging vortices and QPI patterns in $Ca_{2-x}Na_xCuO_2Cl_2$ ($x \sim 0.14$, $T_c \sim 28$ K) at different magnetic fields. All the data were collected with a set-up condition of sample bias voltage $V_s = -100$ mV and tunneling current $I_t = 100$ pA. (**A** to **C**) Vortices imaged by mapping a function $s(\mathbf{r},E) \equiv dg(\mathbf{r},+E)/dV / g(\mathbf{r},+E) - dg(\mathbf{r},-E)/dV / g(\mathbf{r},-E)$ at $E = 4.4$ meV. If there is a gap in the spectrum $g(\mathbf{r}, E)$, the function $s(\mathbf{r}, E)$ below the gap energy takes larger value as gap structure becomes deeper, while it is almost zero if $g(\mathbf{r}, E)$ is structureless. Vortices are imaged as shallower-gap regions (smaller $s(\mathbf{r},E)$) shown in brighter color. Broken circles are guides to the eye. (**D** to **F**) Real-space QPI patterns at $E = 4.4$ meV imaged by mapping the conductance-ratio $Z(\mathbf{r},E)$. (**G** to **I**) $|Z(\mathbf{q},E)|$ obtained by Fourier transforming $Z(\mathbf{r},E)$ shown in (D) to (F). In order to enhance the signal-to-noise ratio, each $|Z(\mathbf{q},E)|$ map is averaged by folding it so as to superpose all the crystallographically equivalent $\mathbf{q}$ positions. Arrows in (G) correspond to those in Fig. 1A.

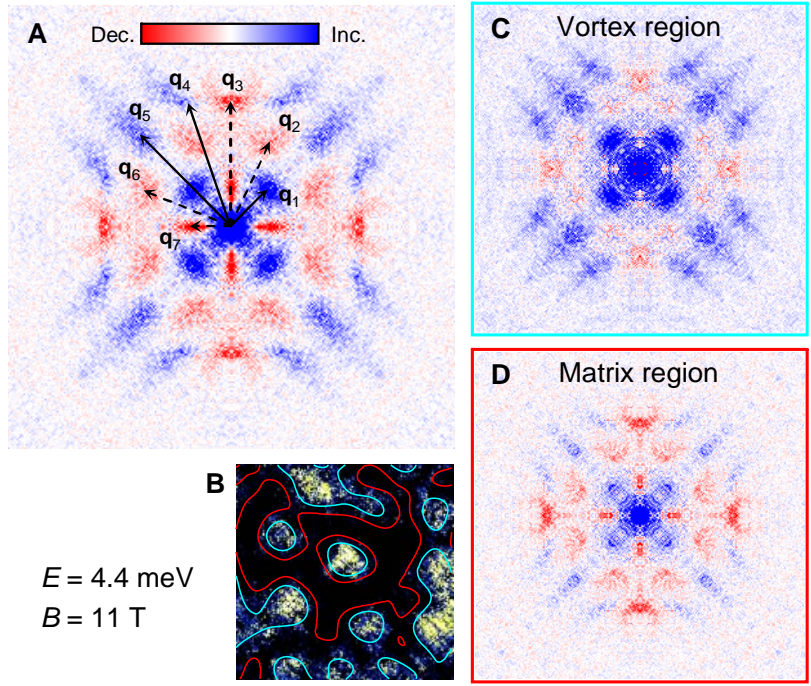

**Fig. 3.** Magnetic-field-induced weight transfer in $|Z(\mathbf{q},E)|$ at $E$ = 4.4 meV. (**A**) The difference map $|Z(\mathbf{q},E,B)|$ - $|Z(\mathbf{q},E, B = 0)|$ for $B$ = 11 T (Namely, difference between Fig. 2I and Fig. 2G). Intensities of sign-preserving **q**-points are field-enhanced while those of sign-reversing ones are field-suppressed. (**B**) Vortex image reproduced from Fig. 1C showing the restricted field of views. Blue and Red lines surround vortex and matrix regions, respectively. (See Fig. 6.) Magnetic-field-induced weight transfers are deduced separately for vortex and matrix regions as shown in (**C**) and (**D**), respectively. Intensities are normalized according to the area. Enhancement of sign-preserving scatterings at $\mathbf{q}_1$, $\mathbf{q}_4$, and $\mathbf{q}_5$ is remarkable near the vortices while it is weak in the matrix region.

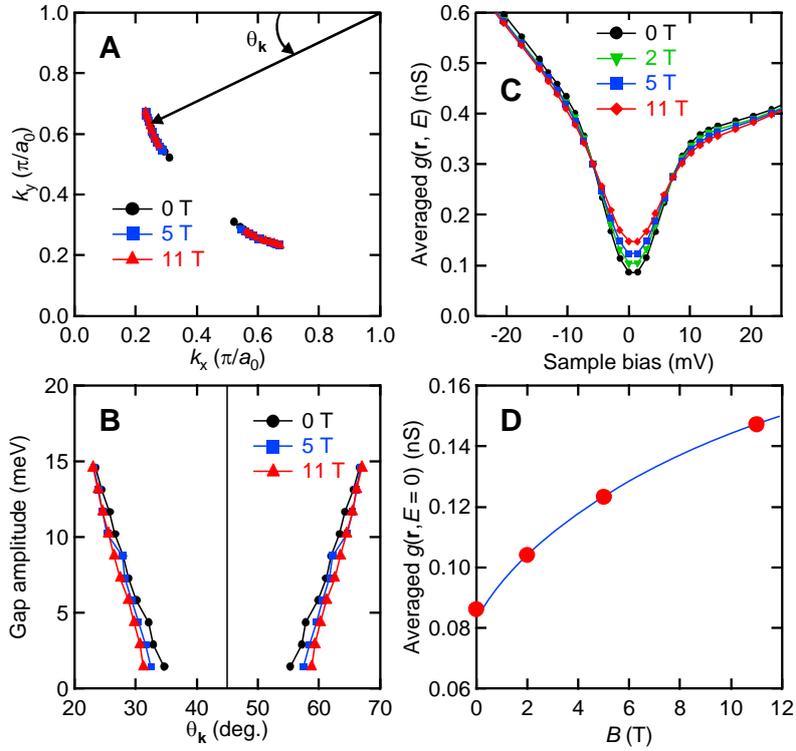

**Fig. 4.** Magnetic-field $B$ effects on the electronic states of in $Ca_{2-x}Na_xCuO_2Cl_2$ ($x \sim 0.14$, $T_c \sim 28$ K). (**A**) Loci of octet ends of constant energy contours at different $B$ representing the underlying Fermi surface. Four independent $q_4(E) = (\pm 2k_x(E), 2k_y(E))$, $(2k_y(E), \pm 2k_x(E))$ were used for analysis. No measurable $B$-induced change is found in the Fermi surface. (**B**) $B$-induced renormalization of the $d$-wave SC gap dispersion. $B$ enlarges the apparent gapless region around the gap node, while the dispersion at higher energy is relatively insensitive to $B$. (**C**) Tunneling spectra averaged over the field of view. Gap-like feature below about 10 meV gets shallower and DOS at Fermi energy ($E = 0$) increases with increasing $B$. Spectrum at 2 T was averaged in the slightly different field of view for other fields. (**D**) The $B$ dependence of spatially averaged $g(\mathbf{r}, E = 0)$. Blue line denotes $B\log B$ behavior expected in a dirty $d$-wave superconductor.

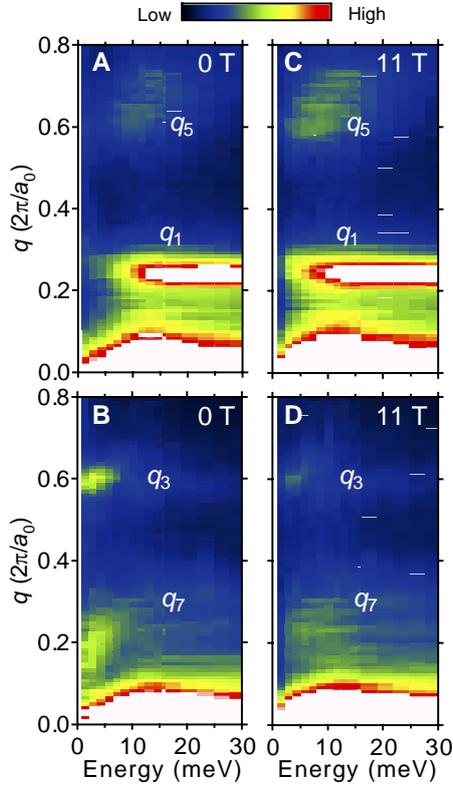

**Fig. 5.** Energy dependence of QPI intensities obtained by taking linecuts from $|Z(\mathbf{q},E)|$ at various $E$ along specific $\mathbf{q}$ directions. Color scale is set to the same range for Figs. 2, G to I. (**A** and **B**) Linecuts along $(0,0)$-$(0,2\pi/a_0)$ and $(0,0)$-$(2\pi/a_0,2\pi/a_0)$, respectively, at $B = 0$ T. Dispersing **q**-vectors are seen below 10 ~ 15 meV. Although intense $\mathbf{q}_1$ peak still exists above this energy, the peak stops dispersing. Other **q** peaks diminishes. (**C** and **D**) Same linecuts at $B = 11$ T. It is clear that intensities of $\mathbf{q}_1$ and $\mathbf{q}_5$ peaks are enhanced while those of $\mathbf{q}_3$ and $\mathbf{q}_7$ peaks are suppressed. Note that $B$-enhanced signals show clear energy dispersion.

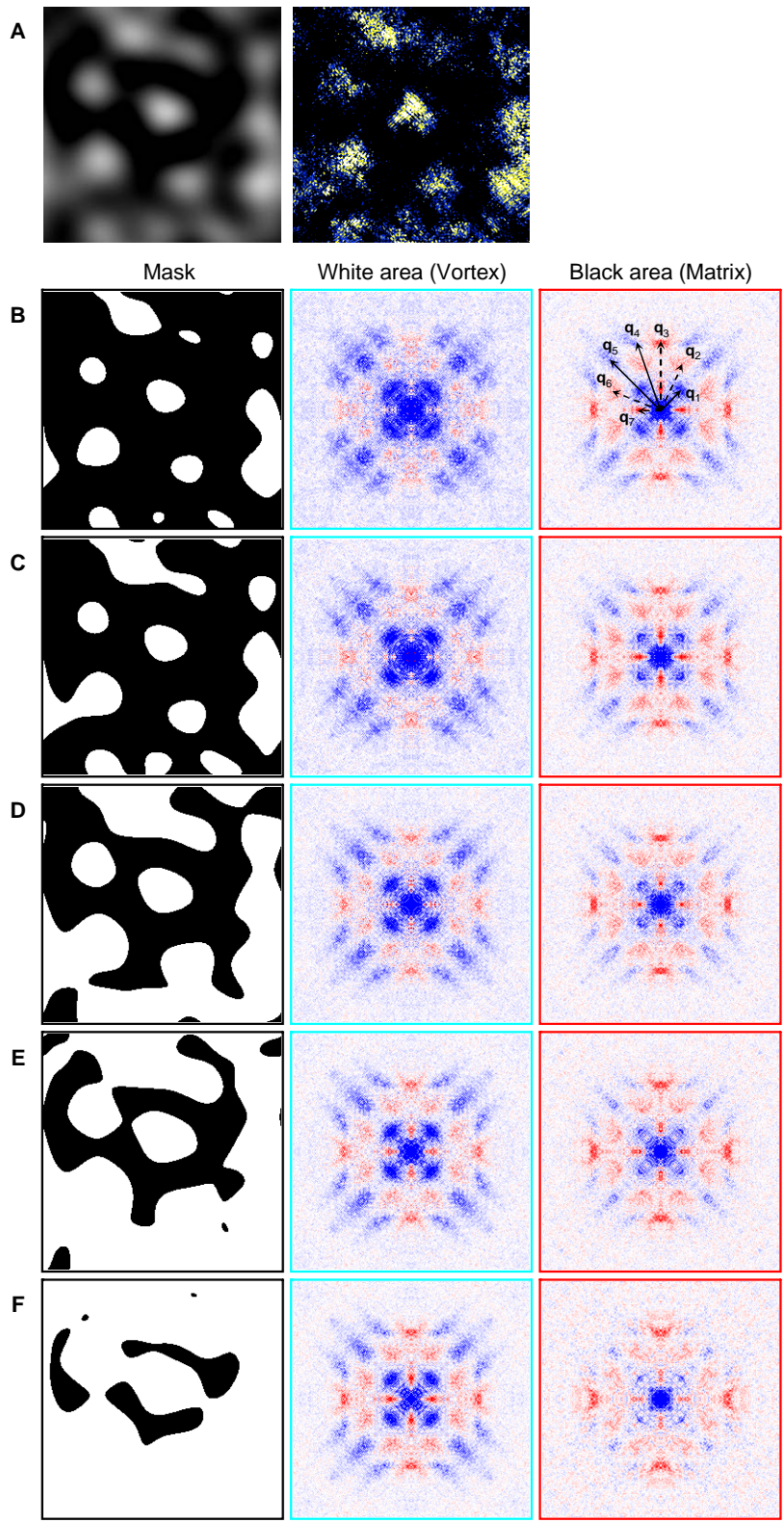

**Fig. 6.** Procedure of the restricted-field-of-view $|Z(\mathbf{q},E)|$ analysis at $E = 4.4$ meV and $B = 11$ T. (**A**) In order to separate vortex and matrix regions, original vortex image (right panel, reproduced from Fig. 2C) is Fourier filtered as shown in the left panel. A circular region around the origin in $\mathbf{q}$ space with a diameter of $0.08 \times 2\pi/a_0$ was used for filtering. (**B** to **F**) Taking contours from the filtered image, a series of masks shown in the left panels of (B) to (F) can be generated. White and black regions denote vortex and matrix regions according to different criteria. In these regions, $|Z(\mathbf{q},E,B)| - |Z(\mathbf{q},E,B = 0)|$ are calculated as shown in central and right panels. Intensities are normalized according to the areas of the restricted fields of view. As indicated in the central column, enhancement of sign-preserving scatterings at $\mathbf{q}_1$, $\mathbf{q}_4$, and $\mathbf{q}_5$ gradually grow as the field of view is restricted to the vortex centers. When the field of view is restricted in the region away from vortices (right column), enhancement of sign-preserving scatterings almost disappears. Figs. 3C and D of the text correspond to the central panel of (C) and right panel of (E), respectively.